# Detection of Shilling Attack Based on T-distribution on the Dynamic Time Intervals in Recommendation Systems

Wanqiao Yuan, Yingyuan Xiao*, Xu Jiao, Wenguang Zheng, Zihao Ling
*Tianjin Key Laboratory of Intelligence Computing and Novel Software Technology, Tianjin University of Technology, Tianjin 300384, China*
*Key Laboratory of Computer Vision and System, Ministry of Education, Tianjin University of Technology, Tianjin 300384, China*
739902931@qq.com, yyxiao@tjut.edu.cn, jiaoxu1999@sina.com, wenguangz@tjut.edu.cn, zihaoling@hotmail.com

*Abstract*—With the development of information technology and the Internet, recommendation systems have become an important means to solve the problem of information overload. However, recommendation system is greatly fragile as it relies heavily on behavior data of users, which makes it very easy for a host of malicious merchants to inject shilling attacks in order to manipulate the recommendation results. Some papers on shilling attack have proposed the detection methods, whether based on false user profiles or abnormal items, but their detection rate, false alarm rate, universality, and time overhead need to be further improved. In this paper, we propose a new item anomaly detection method, through T-distribution technology based on Dynamic Time Intervals. First of all, based on the characteristics of shilling attack quickness (Attackers inject a large number of fake profiles in a short period in order to save costs), we use dynamic time interval method to divide the rating history of item into multiple time windows. Then, we use the T-distribution to detect the exception windows. By conducting extensive experiments on a dataset that accords with real-life situations and comparing it to currently outstanding methods, our proposed approach has a higher detection rate, lower false alarm rate and smaller time overhead to the different attack models and filler sizes.

## I. Introduction

So far, the recommendation systems have been widely used on various e-commerce websites, such as the Amazon personalized product recommendation, Google Reader's individuation reading and all kinds of personalized advertising, etc. The recommendation systems [1] can contact the user and information. On the one hand, it helps users find information that they're interested in. On the other hand, it enables information to be displayed in front of users who are interested in this information. Recommendation systems help users get rid of the trouble of information overload, but it heavily depends on users' historical records, rating records, and other external historical behaviors, etc. Some malicious merchants pour into a large amount of biased user rating profiles in the recommendation systems for their own benefits [2-6], with the purpose of changing the recommending results for their own profits. As a result of, the recommendation results become inaccurate, which adversely affect the reputation of the recommendation systems. Above user rating profile is called shilling attack profile, and this malicious businessman is called a shilling attacker.

The growing number of businessmen has been disturbing the recommendation results. And they get great results with very little effort. This seriously affects the stability and accuracy of the recommendation systems. To solve the problem, Zhang et al. [15] proposed constructing a time series of rating in order to detect abnormal profiles according to sample average and sample entropy in each window. Additionally, Gao et al. propose two abnormal profiles detection methods, the former is based on fixed time intervals, and the latter adopts a dynamic partitioning method for time series. Although they have made progress, the satisfaction of detection rate, false alarm rate and time overhead remains challenging.

In order to solve this issue, this paper proposes T-distribution [8] on the Dynamic Time Intervals, which can solve these problems well through lots of experiments. We summarize the contributions as follows:

- We use the dynamic time intervals method that takes advantage of the quickness of the shilling attack to effectively divide the attack profiles into one window.
- We introduce T-distribution method whose characteristic is the smaller sample and stronger detection ability in order to accurately detect abnormal windows.

The rest sections of the paper are distributed as follows. The second part introduces the common attack models and traditional detection methods. In the third part, we mainly introduce the background knowledge of T-distribution and some problem definitions to facilitate our system. In the fourth part, we detail the method proposed in this paper. And in the fifth part, as the experimental part, we give the evaluation of MovieLens dataset. Finally, we summarize in the sixth part.

## II. Related work

The shilling attack [7] main includes push attacks and nuke attacks. The push attacks can enable the recommendation systems to make it easier to recommend target items, and the nuke attacks are to make it more difficult to recommend target items. To avoid being detected, many attack models have been introduced to disguise themselves [2]. Several common attack models are proposed here. Attack profiles contain filler items, selected items, unrated items and target item(s). Selected items are based on specific needs of the spam user, filler items are a set of randomly selected items to fake normal users, unrated items are those items with no ratings in the profiles, and target items are the items that attackers attempt to promote or demote. The selected items and filler items strengthened the power of the shilling attack, and the attack profile is shown in the Fig. 1.

---

* Corresponding author.
  E-mail addresses: yyxiao@tjut.edu.cn

These basic elements in the Fig. 1. make up different types of attack models. The three most common attack models are random attack, average attack, and popular attack. Random attack is to randomly select some items from the system for random ratings. Its advantage is that requires little cost but gains great benefits. Average attack is that items of random attack are assigned the corresponding average ratings, and in the literature [5], Burke et al. find this attack model win the advantages over the random attack. Popular attack is an extension of random attack, on this basis, the most popular item in the field is selected to have the highest rating. The model structure is shown in the table I. In addition, in order to avoid detection, the attacker also adds obfuscated attack, which is more difficult to detect [10,11].

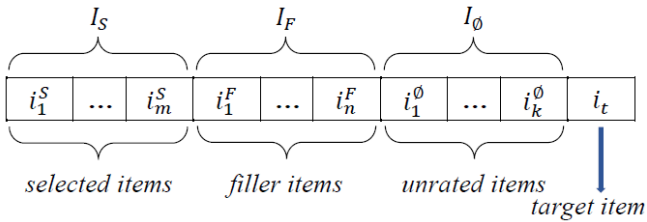

Fig. 1. The general of form of a shilling attack profile

TABLE I. THREE CLASSICAL ATTACK MODELS

| Attack Type | Push Attack |
|---|---|
| Random | $I_S = \Phi, I_F = r_{ran}, I_t = r_{max}$ |
| Average | $I_S = \Phi, I_F = r_{avg}, I_t = r_{max}$ |
| Bandwagon | $I_S = r_{max}, I_F = r_{ran}, I_t = r_{max}$ |
| **Attack Type** | **Nuke Attack** |
| Random | $I_S = \Phi, I_F = I_{ran}, I_t = r_{min}$ |
| Average | $I_S = \Phi, I_F = I_{ran}, I_t = r_{min}$ |
| Bandwagon | $I_S = r_{max}, I_F = r_{ran}, I_t = r_{min}$ |

The early detection methods include supervised learning and unsupervised learning technologies [12, 13]. No matter where there are methods, they are all aimed at analyzing the features of user profiles. For examples, Williams et al. [13] proposed a kind of supervised learning method through extracting a series of the user profile characteristics, including Rating Deviation from Mean Agreement (RDMA), Length Variance (LengthVar), Degree of Similarity with Top Neighbors (DegSim), and then examine attack profiles by the supervised learning algorithm. However, this method only makes effort to a specific type of attack model and has not universality. Bryan et al. [14] proposed an Unsupervised Retrieval of Attack Profiles algorithm (UnRAP), which found attacks could readily be identified by biclustering, so an unsupervised algorithm of introducing Hv-score values is proposed to detect Attack profiles. This method wins the advantages to be able to detect different Attack model. Wu et al.[17] put forward the semi-supervised learning Attack detection algorithm (Semi-SAD), which combines supervised learning with unsupervised learning and combines Naïve Bayes with EM-$\lambda$ algorithm to detect attack profiles by optimizing parameters $\lambda$. But all of the above methods have high calculation cost and don't have good detection rates.

Zhang et al. [15] and Gao et al. [9, 16] analyzed the changes in the target item, which reduced the calculation engineering and improved the universality and accuracy. Zhang et al. proposed a time series detection method based on the sample mean and sample entropy according to the rapidity and purpose of the attacker. However, the time window of this method is fixed, and different time Window sizes have diverse detection effect. Gao et al. proposed dynamic partitioning for time series based on significant points, and then used $\chi^2$ to identify item-rating abnormal time window. However, $\chi^2$ heavily depended on the population variance, and with dynamically divided time intervals, every item-rating sample of time intervals has different sparsity and density. Although the more samples there are, the more obvious the detection effect will be, however, in the time window of small samples, the detection efficiency is often not obvious. Therefore, we propose a T-distribution detection based on the Dynamic Time Intervals method to solve this problem more accurately.

III. PRELIMINARIES

T-distribution was first proposed by William Sealy Gosset in 1908. T-distribution is dramatically significant in Statistics researches, in order to tackle several practical problems through Interval Estimation and Hypothesis Testing [18], which has significant applications in the medical domain to detect blood quality problems. The shape of the T-distribution curve is related to the degree of freedom $v$, which is related to the sample size $n$. Compared with the standard normal distribution curve, the smaller the number of samples is, the smaller the degree of freedom $v$ is, the flatter the T-distribution curve is, the lower the middle of the curve is, and the higher the tail of the curve is, oppositely, the higher the degree of the sample is, the closer the T-distribution curve is to the positive distribution curve. When the degree of freedom $v$ is infinite, the curve of the T-distribution is the standard positive distribution curve. That is to say, This feature of the T-distribution is significant for us to detect samples injected with a small number of attacks.

In order to facilitate our system, we have the following notations and definitions:

- $I$: the set of the entire items.
- $U$: the set of the entire users.
- $H'$: the set of the rating actions. Specifically, each rating action $h \in H'$ is represented as $h = <h.i, h.u, h.r, h.t>$, where $h.i \in I$ refers to an item, $h.u \in U$ denotes the user that gives $h.i$ a rating, $h.r$ is the rating, and $h.t$ is the time of rating.

**Definition 1 (Rating History).** For each item $i_k \in I$, a rating history $H_k$ of item $i_k$ is a sequence of rating records formatted as $H_k = h_1 \xrightarrow{h_2.t-h_1.t} h_2 \xrightarrow{h_3.t-h_2.t} ... \xrightarrow{h_n.t-h_{n-1}.t} h_n$, where $h_j \in H'$, $0 < j < n$, $h_{j+1}.t > h_j.t$, and $\nexists h_{j'} \in H'$, s.t. $h_j.t < h_{j'}.t < h_{j+1}.t$.

**Definition 2 (Item-ratings Time Gaps Series).** For each rating history $H_k$, a item-ratings time gaps series is $IRTGS_k = \{(midT_1, gap_1), ...,(midT_{n-1}, gap_{n-1})\}$, $midT_j = (h_{j+1}.t - h_j.t)/2$, refers to the median of the adjacent timestamps between ratings, and $gap_j = h_{j+1}.t - h_j.t$, refers to the adjacent timestamps gap between ratings.

**Definition 3 (Time Window).** Suppose $H_k$ is divided into m time intervals, time window series of $H_k$ is $W_k = \{w_1, w_2, ..., w_m\}$, each time window $w_x \in W_k$ corresponds to all $h \in H_k$ of the $x$th time interval. Besides, $w_1 \cup w_2 \cup ... \cup w_m = H_k$, $w_1 \cap w_2 \cap ... \cap w_m = \emptyset$.

**Definition 4 (Window Size).** For each time window $w_x \in W_k$, a window size of $w_x$ is $ws_x$ that refers to the amount of containing all $h$ in $w_x$. For example, Suppose $h_j \in H_k$ and $w_x = \{h_j, h_{j+1}, ..., h_{j+z}\}$, window size of $w_x : ws_x = z$.

## IV. T-DISTRIBUTION TO DETECT ATTACKS ON THE DYNAMIC TIME INTERVALS METHOD

In order to gain more benefits, attackers inject a large number of high ratings (push attack) and low ratings (nuke attack) into the target item in a fast time to make the recommendation system easier or harder to recommend the item. Therefore, a common characteristic of all attack models is that a large number of attack profiles will be injected into the target item in a small time interval. That is to say, high ratings or low ratings and the number of rating will be significantly increased or drastically decreased in the attack time interval. According to the features of shilling attack mentioned above, we propose TDTI (T-distribution to Detect Attacks on the Dynamic Time Intervals) method.

Our method is mainly based on two features of the attack as preconditions: 1) Attackers inject a large number of fake profiles in a short period in order to save costs, and attacks must be very dense. 2) Incidence rate of injection periods must be tiny time intervals throughout entire lifecycle of the item.

The general idea of the proposed method is as follows: Firstly, we divide the rating history of item into time windows by using the algorithm of DTI (Dynamic Time Intervals). And then, through the T-distribution algorithm, we calculate the T-value between each time window and other time windows of the rating history in order to determine suspicious time windows that are different from others in the rating history. To be more precise, through analyzed the timestamp gaps of time windows, we obtain abnormal windows. Finally, we calculate the mean of ratings in the abnormal window, exclude the ratings less than the mean (push) or the ratings more than the mean (nuke), and the rest is the abnormal ratings of the attacks.

In this section, we will introduce our TDTI method in detail, divided into two modules: 1) designing the DTI algorithm partitioning rating history of item into time window series. 2) T-distribution algorithm identifies abnormal windows.

### A. Designing the DTI algorithm partitioning rating history.

We aim to partition the rating history of item into multiple time windows, and ensure that the attack profiles are divided into one time window in the same period. Based on feature1: a large number of attacks are injected system in a short period, the specific steps are as follows:

*1)* According to the rating history of item, we calculate the corresponding IRTGS (Item-ratings Time Gaps Series).

*2)* Divide IRTGS into two subsequences according to the maximum value of $gap$ and the correspond $midT$.

*3)* Repeat step2 for the two subsequences again, and record the $midT$ value as mark point, until the difference between maximum and minimum of $Gap$ is less than $\alpha$. In order to make the implementation efficiency high, we set $\beta$ value, if the length of IRTGS is less than $\beta$ value, the circulation ends. This is mainly because of the rapidity of the shilling attack, with very intensive rating at very small intervals, so the $\alpha$ value cannot be too large, otherwise the normal rating will be integrated into the abnormal window with a large number of attacks. Meanwhile also cannot be too small, otherwise, the attack profiles will be divided into multiple windows. The $\alpha$ value is related to the rating characteristic, and $\alpha$ value is obtained by a large number of experimental tests.

*4)* Divide the rating history of item based on the mark points recorded by step3.

Algorithm 1 describes the process of DTI. Lines 1 and 2 correspond to step1, and lines 3 to 12 correspond to step2 and step3. Lines 12 and 13 correspond to step4.

---

**Algorithm 1**: DTI partitions rating history

**Input:** the rating history of item: $H\{h_1, ..., h_i, ..., h_n\}$, threshold values $\alpha$ and $\beta$
**Output:** all time windows of item $W\{w_1, ..., w_m\}$
1: **for** $i$ **from** 0 **to** $n-1$:
2:   get $gap_i, midT_i$
3: Constructing **dti** $(gap, midT)$:
4:   **if** $(gap.\text{length} > \beta$ **and**
         $gap.\max - gap.\min > \alpha)$:
5:     $m = gap.\max.\text{location}.$
6:     $gap\_left = gap[0:m]$
7:     $gap\_right = gap[m+1:gap.\text{length}]$
8:     $midT$ does the same thing
9:     $mark.\text{append}(midT[m])$
10:    **dti**$(gap\_left, midT\_left)$
11:    **dti**$(gap\_right, midT\_right)$
12: sort $(mark)$ group $H$
13: get $W$

---

TABLE II. THE FREEDOM CORRESPONDS TO BOUNDARY VALUE IN THE 95% CONFIDENCE LEVEL

| Freedom | 1 | 2 | 3 | 4 | 5 | 6 | 7 | 8 |
|---|---|---|---|---|---|---|---|---|
| **Boundary value** | 12.71 | 4.303 | 3.182 | 2.776 | 2.571 | 2.447 | 2.365 | 2.306 |

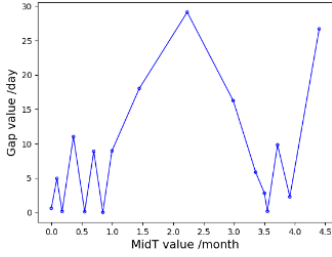 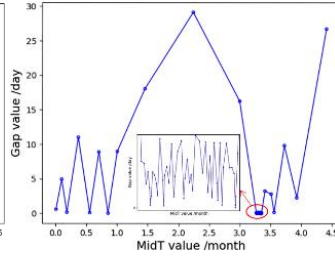 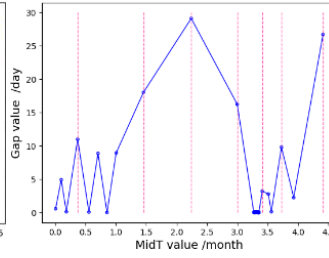 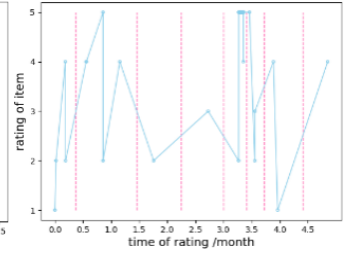

| Fig. 2. IRTGS of Non-Attacks | Fig. 3. IRTGS of Attacks | Fig. 4. recording point marks | Fig. 5. cutting rating history |

The above step is the DTI to divide the rating history into multiple time windows. For example, In order to illustrate our method more clearly, we randomly selected an item from MovieLens 100k dataset. The IRTGS of original data of the item is shown in Fig. 2. A large number of attack profiles are injected in a short time in Fig. 3 and here we injected 50 attack profiles to show clear result, where a host of attack profiles are shown in the red circle mark in which the smaller image in the middle of Fig. 3 is zoomed. Fig. 4 shows the mark points are recorded by the method of dynamically dividing IRTGS. We record the horizontal coordinates corresponding to the red vertical line as mark point set in order to divide the rating history. Fig. 5 shows the cutting of rating history based on the mark point. Obviously, according to mark point set divides the rating history into time window series, and attacks are divided into a time window, which contains tiny normal profiles.

### B. T-distribution algorithm Identifies Abnormal Windows.

After The abnormal windows are detected by the following steps:

*1)* Calculate the T value of each time window and other Windows to determine whether each window is consistent with the properties of others.

*2)* Compare the relationship between each T value and its corresponding boundary value. If the value beyond the boundary value is denoted as 1 (inconsistent), and if it is denoted as 0 within the boundary value range (consistent). Sequentially, we obtain a $time\ window - time\ window$ Symmetric Matrix, where its values only have 0 or 1, so in this paper, we call it $0 - 1$ Matrix.

*3)* Analyze the $0 - 1$ Matrix, and we select the windows whose 1 value quantities are more than the mean of 1 value quantities of all windows in rating history to get suspicious window set. Based on the shilling attack characteristics of 1 (Attackers inject a host of false profiles in a short period) and 2(Incidence rate of injection periods must be tiny time intervals throughout entire lifecycle of the item.), abnormal window must be a minority, and the case that the windows get 1 value is more than the normal windows, Thus, we can easily get suspicious windows by comparing the number of 1 value in each window to the mean.

*4)* Compare each window in the suspicious window set obtained in step 3, if time gap of the window is greater than the mean time gaps of all time windows in the rating history and the rating number of the window is smaller than the mean rating quantities of all Windows in the rating history. Based on the attack characteristics of 1 (rapidity), we reasoned that each timestamp gap of abnormal window is smaller than timestamp gaps of normal windows and each rating quantity of abnormal window is more than normal windows. Through this condition, we get the attack windows.

The T value was calculated by the modified two-sample T-distribution hypothesis testing method[18]. The formula applied to the procedures is as follows:

$$\bar{x}_i^* = \frac{1}{m}\sum_{k=1}^{m} x_{ik} \quad (1 \le m \le 5) \qquad (1)$$

$$\bar{x}_j^* = \frac{1}{n}\sum_{k=1}^{n} x_{jk} \quad (1 \le n \le 5) \qquad (2)$$

The formula (1) and (2) are functions of the means in time window. $\bar{x}_i^*$ is the modified mean of the $i$th time window and $\bar{x}_j^*$ represents modified mean of the $j$th time window. Since the score range is from 1 to 5 in this context, the max rating difference is 5, leading to that the different between the rating mean of time windows can only be within the range of 5 and the calculation of T value is related to the difference, which is no obvious effect. In order to increase the difference, we improve the function through the $m$ in formula (1) and $n$ in formula (2), refer to the number of rating kinds rather than the number of ratings, so that we can realize the more the rating number and the larger the difference change of rating mean. $x_{i_k}$ in formula (1) refers to the $k$th rating in the $i$ window and $x_{j_k}$ in formula (2) is the $k$th rating in the $j$ window.

$$\bar{x}_i = \frac{m}{g} \bar{x}_i^* \qquad (3)$$

$$\bar{x}_j = \frac{n}{h} \bar{x}_j^* \qquad (4)$$

The formula (3) is conversion function of modified mean ($\bar{x}_i^*$) and traditional mean ($\bar{x}_i$), and formula (4) shows relationship between modified mean ($\bar{x}_j^*$) and traditional mean ($\bar{x}_j$), where $g$ and $h$ refers to the number of rating in the $i$th and $j$th time windows.

$$s_i^2 = \frac{1}{g}\sum_{k=1}^{g}(x_{i_k} - \bar{x}_i)^2 \qquad (5)$$

$$s_j^2 = \frac{1}{h}\sum_{k=1}^{h}(x_{j_k} - \bar{x}_j)^2 \qquad (6)$$

The formula (5) and (6) are variances of time window.

$$T_{ij} = \frac{\bar{x}_i^* - \bar{x}_j^* - (a_0 - a_i)}{\sqrt{gs_i^2 + hs_j^2}} \sqrt{\frac{mn(m+n-2)}{m+n}} \sim t(m+n-2) \quad (7)$$

T value can be expressed as formula (7), where $a_0$ refers to the rating mean in entire rating history, and $a_i$ refers to the rating mean of rating history excluding the $i$th window. $m + n - 2$ is the freedom about T value, which corresponds to the boundary value, shown as Table II. If the freedom is 0, it means that the rating type of both windows is 1, in this case, we default to similar attributes of the two windows.

Algorithm 2 describes the process that T-distribution detects abnormal windows, where lines 1 to 8 correspond to step1, lines 9 to 11 correspond to step2, and lines 12 to 17 correspond to step3 and step4.

Subsequently, we take the item selected in the *A* subsection as an example. Table III shows the value matrix $T_{ij}$ through T-distribution procedure obtained. And then, compare each T value with the corresponding boundary value to obtain the matrix $0 - 1$ in table IV, As can be seen from the table III, the third window far exceeds the boundary value.

TABLE III.    $T_{ij}$ VALUE MATRIX OBTAINED THROUGH OUR PROCEDURE

| $T_{ij}$ | 1 | 2 | 3 | 4 |
|---|---|---|---|---|
| 1 | 0 | 2.534 | 53.013 | 0.157 |
| 2 | 2.683 | 0 | 50.833 | 2.43 |
| 3 | 52.195 | 49.962 | 0 | 52.132 |
| 4 | 0.307 | 2.348 | 53.008 | 0 |

TABLE IV.    $T_{ij}$ VALUE MATRIX OBTAINED THROUGH OUR PROCEDURE

| $T_{ij}$ | 1 | 2 | 3 | 4 |
|---|---|---|---|---|
| 1 | 0 | 0 | 1 | 0 |
| 2 | 0 | 0 | 1 | 0 |
| 3 | 1 | 1 | 0 | 1 |
| 4 | 0 | 0 | 1 | 0 |

## V. EXPERIMENTAL EVALUATION

### A. Datasets

In the experiments, we use public available dataset MovieLens 100k, which has been collected by GroupLens Research and made available from the GroupLens web site. MovieLens is a recommendation system; Its main function is to make recommendations based on user preferences and the widely used technology is collaborative filtering. The dataset contains 1682 items, 100000 ratings that were evaluated by 943 users. Each user makes at least 20 ratings, and rating has five ranks from 1 to 5. To facilitate our experiment, we omitted the items with less than 10 rating quantities, because the number of rating was too small to reflect the authenticity of the experiment, and the dataset after filtering is shown in the table V.

TABLE V.    MAINLY INFORMATION OF THE DATASET IN THE EXPERIMENT

| Dataset | Item | User | Rating |
|---|---|---|---|
| MovieLens 100K | 1152 | 943 | 97953 |

**Algorithm 2**: T-distribution identifies attack windows

**Input:** the rating history of item: $H$, all time windows of $H$: $w_1, ..., w_i, ..., w_m$, $w_i\{h_1, ..., h_k\}$, the size of $w_i$: $ws_i$, the number of rating kind of $w_i$: $rk_i$, the timestamp difference of $w_i$: $wd_i$
**Output:** the abnormal time windows $AW$
1: $a_0$ = mean of $H.r$
2: **for** $i$ **from** 0 **to** $m$:
3:    **for** $j$ **from** 0 **to** $m$:
4:      calculate means and variances of $w_i.r$ and $w_j.r$
5:      get $x_i, x_j, s_i^2$ and $s_j^2$
6:      $a_i$ = mean of $H.r$ without $w_i$
7:      **if** ($i! = j$):
8:        $x_i, x_j, s_i^2, s_j^2, a_0, a_i, rk_i, rk_j, ws_i, ws_j$ calculate $t_{ij}$
9:        $rk_i, rk_j$ calculate freedom
10:     compare boundary
11: get matrix $T(0-1)$
12: **for** $i$ **from** 0 **to** $m$:
13:    $z_i = T_i.count(1) - T_j.count(0)$
14: **for** $i$ **from** 0 **to** $m$:
15:    **if** ($z_i >= mean(z)$ **and** $wd_i <= mean(wd)$ and $ws_i >= mean(ws)$):
16:      $AW$.append($w_i$)
17: get $AW$

### B. Evaluation Metrics

We used two indicators [9, 16] to evaluate the experimental results: The detection rate in formula (8) is defined as the number of checked attack profiles divided by the number of the total attack profiles. The false alarm rate in formula (9) is defined as the number of normal profiles that are recognized as abnormal profiles divided by the number of all normal profiles.

$$DetectionRate = \frac{Detected\ Attack\ quantity}{Total\ Attack\ quantity} \quad (8)$$

$$FalseAlarmRate = \frac{False\ quantity}{Normal\ quantity} \quad (9)$$

### C. Comparative Approaches

At present, there are many methods [12,13,17] for attacks detection, but we put forward methods based on time and item view, so we compare it with similar methods [9,15,16] in this paper. The comparison of relevant methods has Zhang et al. [15] and Gao et al. [9,16] in the simulation experiment on the MovieLens 100k dataset. The following are brief introduction to three methods of them:

DP [9]: According to two features of shilling attacks (item abnormality: the rating of item is always maximum and minimum as well as attack promptness: it takes a very short period time to inject attacks), this method first dynamically partitions item-rating time series based on important points, and then, use chi square distribution to detect abnormal intervals.

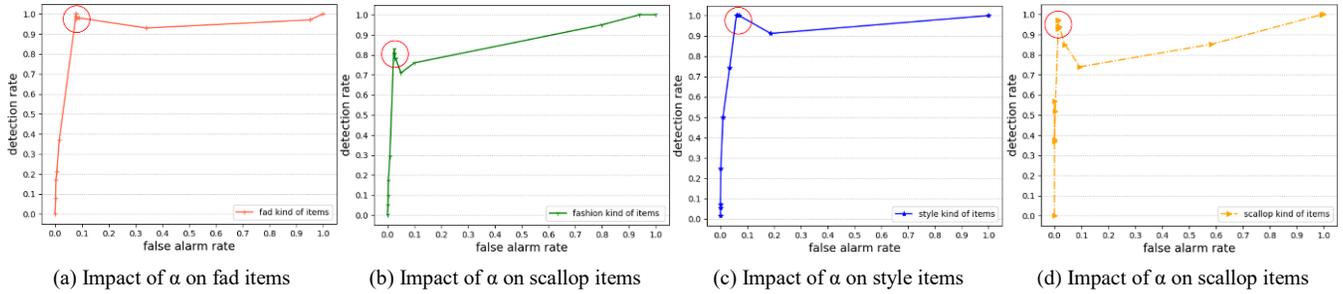

(a) Impact of α on fad items   (b) Impact of α on scallop items   (c) Impact of α on style items   (d) Impact of α on scallop items

Fig. 6. Impact of α(h) on detection and false alarm among the four kinds of items: fad, fashion, style, scallop

TIC [16]: The distributions of ratings in different time intervals are compared to detect anomaly intervals based on the calculation of chi square distribution ($\chi^2$).

TS [15]: The method constructs a time series of rating for an item to compute the features of sample average and sample in each window, and based on duration of attacks, observing the time series of two features can expose attack events, in which sample average is represented by TS-Ave and sample entropy is represented by TS-Ent.

### D. Experiment Performance

To evaluate the performance of our method in different attack quantities, filler sizes, items of feature and attack models, we use MovieLens dataset to take four types ratings: fad, fashion, style, and scallop as well as three attack models: average attack, random attack, and bandwagon attack. Supposing that original users are normal users, and attackers are those generated from attack models. The number of attack is 10, 20, 30, 40, 50 and filler size is set to 1%, 3%, 5%, 7%, 10%, which divided into two categories experience: push and nuke. Based on the characteristics of shilling attack in section (attacks are very dense and injection periods are tiny time intervals in the rating history), we set the time gap between attacks in the same period within 1000 seconds. 50 items are randomly selected from 1152 items. We repeat the experiments fifty times.

This section is mainly divided into three parts to show. Firstly, we introduce the effect of threshold values on the experiment. Secondly, we show the impact of our method compared with other similar experiments. Thirdly, we demonstrate effect of our method to detect four styles of characteristic ratings and three kinds of attack models.

*1) Impact of threshold setting in experimentg:* DTI has two thresholds to be determined in the experiment. One is the $\alpha$ value that controls the end action of DTI. And the other is the $\beta$ value that affects the time efficiency of the experiment.

The selection of $\alpha$ value is related to the four type characteristics ratings. We obtain the value by experiment, and 50 push attacks (filler size is 0) are injected into the system and testing, without thinking about $\beta$ ( $\beta$ is 0). According to literature [16], 1,152 items are divided into four types based on the characteristics of the rating quantity and the rating time, fad, fashion, style, scallop, respectively. The following are the characteristics of four types:

- Fad: There are 75 items in total. It is characterized by relatively dense rating time and fewer rating quantities.
- Fashion: There are 41 items in total. The rating feature is that the rating time is relatively concentrated and the number of ratings is larger.
- Style: There are 250 items in total. It is characterized by scattered rating time and smaller number of rating.
- Scallop: There are 786 items in total. The rating feature is that there is always a topic and the rating quantity is relatively larger.

Fig. 6 shows the effects of threshold α changes on DTI procedure. As shown from Fig. 6, horizontal coordinate is the false alarm rate and vertical coordinate is the detection rate, where one $\alpha$ value corresponds to a detection rate and a false detection rate. Through the increasing change of $\alpha$ value from 0, the broken line graph is drawn, and unit of $\alpha$ is hour (h). In the experiment, we expect the detection rate to be as large as possible and the false alarm rate as small as possible. The most ideal state is that $\alpha$ exists one value, which makes the detection rate is reach 1, and the false alarm rate is 0. That is to say, we need the closest point to the point (0,1). Fig. 6-(a), Fig. 6-(b), Fig. 6-(c) and Fig. 6-(d) are respectively experimental results of fad, fashion, style and scallop kinds of items and we've circled the closest points in red and acquire the ranges of the corresponding $\alpha$ optimal value to structure in table VI, where we list the range of $\alpha$ and the corresponding range of detection rate and false alarm rate in four kinds of items. By integrating the intersecting parts of the four categories items, we get the best experimental results for four kinds of items when the α value is equal to 0.389h, as shown in the synthesize column in table VI.

$\beta$ value is related to the number of attacks injected in the system. According to a large number of experiments, when $\beta$ is set to 10, the experimental result is the best, and the running time of DTI reach the optimal value.

*2) Comparison with Other Approachess:* All compared methods involved Zhang et al.[15] and Gao et al.[9,16] are based on the simulation experiments on the MovieLens 100k *data* set. Since the attack model is not affected in this paper and three article [9, 15, 16], the filler size isn't considered here. The experimental results are shown in Fig. 7.

For detection rate: the experiment results of push attack are shown in Fig. 7-(a).The curve of our method (TDTI) is similar to Gao's method (DP), because they all started with the idea of dynamic partitioning time series. When attack size is 10, the detection rate of TDTI gets to 0.965 and DP is 0.75. When more

TABLE VI. α OPTIMAL VALUE IN DIFFERENT KINGDS OF ITEM (α UNIT: H )

| kinds of item | fad | fashion | style | scallop | synthesize |
|---|---|---|---|---|---|
| α | 0.278-0.389 | 0.389-0.444 | 0.278-2.78 | 0.278-0.389 | 0.389 |
| detection rate | 1 | 0.82-0.84 | 1 | 0.96-0.97 | 0.979 |
| false alarm rate | 0.0708-0.072 | 0.023-0.025 | 0.062-0.063 | 0.015 | 0.028 |

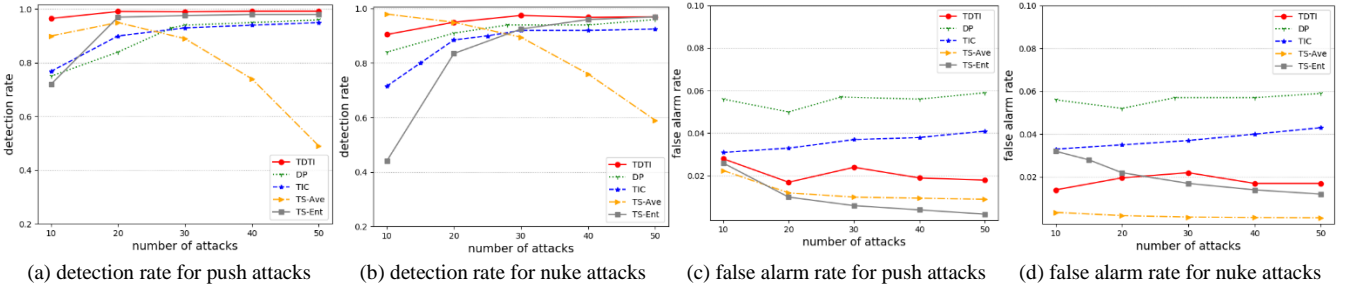

(a) detection rate for push attacks  (b) detection rate for nuke attacks  (c) false alarm rate for push attacks  (d) false alarm rate for nuke attacks

Fig. 7. Detection rates and false alarm rate on MovieLens 100k data set

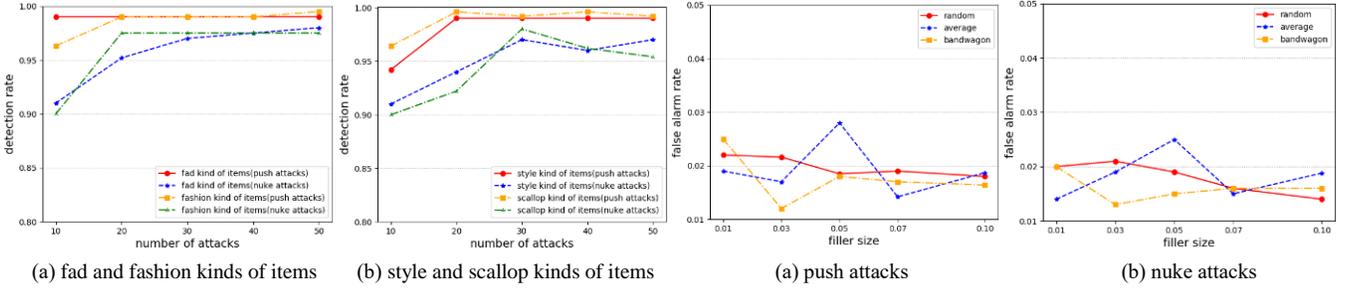

(a) fad and fashion kinds of items  (b) style and scallop kinds of items

Fig. 8. Impact of attack size on four kinds of items

(a) push attacks  (b) nuke attacks

Fig. 9. Impact of filler size on attack model

than 20, TDTI maintains steady state, slowly rising from 0.99 to 0.992, and the DP is also rising from 0.93 to 0.96. It is obvious from the data that our method is more accurate, because DTI method can more accurately dividing the attack profile at a time window than Gao's method, and for the detection of unusual window, we make full use of the characteristic of T-distribution where small sample detection effect is obvious. Zhang's method (TS-Ent) has a very bad rate that is 0.72 when attack size is 10. TIC doesn't have obviously better effect than the DP and TDTI methods.

Fig. 7-(b) shows the experiment effect of nuke attack. When Attack size at 10, TDTI is 0.905, and Zhang's method (TS-Ave) is 0.98, which is 0.075 higher than ours. However, the curve of sample average declines linearly, When the attack size is 50, the detection rate of his method is 0.59. This is that the method used by Zhang is static window, so the larger the attack is, the more easily the attack is divided into multiple windows, resulting in decreasing of detection rate. TDTI curve is relatively stable, when attack size is 50, the detection rate was 0.97.

For false alarm rate: as shown in Fig. 7-(c) and Fig. 7-(d), the rate of Gao's method is very high, while Zhang's method is very low. The rate of our method (TDTI) is above Zhang's method and below the Gao's method, and the trend is relatively stable in the below 0.028. This illustrates that T-distribution method has an obvious identification effect on determining the similarity between windows.

Through the comparison between methods in the experiment, our method has a satisfactory effect, which is because we clearly know the common characteristics of the shilling attack, and put forward the detection method according to its characteristics and existing conditions.

*3) Comparison of between rating features and between attack models:* We detect the effect of our method in the four kinds of items that fad, fashion, style and scallop, and in the three kinds of common attack models which including random attack, average attack and bandwagon attack. As well as we observe the change of detecting results by changing the attack size and filler size.

The following Fig. 8 is the evaluation of the detection rate of the four types items, in the average attack condition that the injection attack size is 10 and the filler size is 5%. The detection rate of push attacks are generally higher than nuke attacks. When the attack size is 10, the difference between push and nuke is the largest, where the detection rate of fad item injecting push attacks is 0.99 and nuke is 0.91, which is 0.07 higher. The best experiment effect of four types is fad, whose push attack gets to 0.99 as well as nuke attack is more than 0.91, and the highest is up to 0.98. This is because the fad rating features are time-intensive and less quantity, which makes it easier for our method to divide normal profiles into a time window and attack profiles into a time window, through the T-distribution method can more accurately compare the similarity of normal and abnormal window. Among them, scallop has the worst effect compared with the other three types. The reason for the low detection rate is that scallop has ratings in each time period and the number of ratings is large, which makes it difficult to divide attack profiles into a time window where there are less normal profiles. In the experiment, there is little difference in false alarm rate between the four rating types, as they have the same condition that filler size is 10%.

Fig. 9 is to test the influence of the filling size on each attack model. Experimental conditions: the attack size is 10, and the selected item in the Bandwagon attack is 50th, because the item always has the topic degree and its rating quantity is largest in the 1682 items.

As shown in the Fig. 9, the false alarm rate of the three types is below 0.028, and the lowest is 0.012. The false alarm rate of the attack model did not increase with the increasing filling size. The highest of these is the average attack. Random attack and popular attack are stable. Since many articles [9,15,16] have demonstrated that the detection methods based on item are not greatly affected by the attack model, we no longer consider the influence of the attack model on the detection rate.

## VI. CONCLUSIONS

Based on the characteristics of the shilling attacks, this paper proposes the DTI algorithm to divide the rating history into multiple time windows, and then the T-distribution algorithm calculate the similarity between windows, and finally, system identifies shilling attacks by analyzing the time difference and profile quantity of each window.

Compared to other methods, detection rate of our method achieves the desired value, but our false alarm rate still needs to be improved. Therefore, we should continue to work hard in order to reduce the false alarm rate. Our future work is striving to make better results.

ACKNOWLEDGMENT

This work is supported by the National Nature Science Foundation of China (61702368), Major Research Project of National Nature Science Foundation of China (91646117) and Natural Science Foundation of Tianjin (17JCYBJC15200, 18JCQNJC0070)

REFERENCES

[1] D. H. Park, H. K. Kim, I. Y. Choi, and J. K. Kim, "A literature review and classification of recommender systems research," International Conference on Social Science and Humanity, vol. 5, 2011, pp. 10059-10072.

[2] R. Burke, B. Mobasher, C. Williams, and R. Bhaumik, "Classification features for attack detection in collaborative recommender systems," in ACM SIGKDD International Conference on Knowledge Discovery and Data Mining, 2006, pp. 542-547.

[3] M. P. O'Mahony, N. Hurley, N. Kushmerick, and G. Silvestre, "Collaborative recommendation: A robustness analysis," in ACM Transactions on Internet Technology, vol. 4, no. 4, November 2004, pp. 344-377.

[4] S. K. Lam and J. Riedl, "Shilling recommender systems for fun and profit," in ACM International Conference on World Wide Web, WWW 2004, New York, USA, May 2004, pp. 393-402.

[5] M. P. O'Mahony, N. J. Hurley, and G. C. M. Silvestre, "Recommender Systems: Attack Types and Strategies," Twentieth National Conference on Artificial Intelligence and the Seventeenth Innovative Applications of Artificial Intelligence Conference, 2005, pp. 334-339.

[6] T. Zhou, Z. Kuscsik, J. G. Liu, and M. Medo, "Solving the apparent diversity-accuracy dilemma of recommender systems," Proceedings of the National Academy of Sciences of the United States of America, 2010, pp. 4511-4515.

[7] B. Mehta, T. Hofmann, and P. Fankhauser, "Lies and propaganda: detecting spam users in collaborative filtering," in ACM International Conference on Intelligent User Interfaces, IUI 2007, Honolulu, Hawaii, USA, January 2007, pp. 14-21.

[8] E. Grosswald, "The Student t-Distribution for Any Degrees of Freedom is Infinitely Divisible," The Annals of Probability, Springer-Verlag, 1976, pp. 680-683.

[9] M. Gao, et al., "Item Anomaly Detection Based on Dynamic Partition for Time Series in Recommender Systems," PLOS ONE, vol. 10, 2015, pp. 135-155.

[10] Z. Cheng and N. Hurley, "Effective diverse and obfuscated attacks on model-based recommender systems," in ACM Conference on Recommender Systems, RecSys 2009, New York, Ny, Usa, October 2009, pp. 141-148.

[11] R. Burke, B. Mobasher, C. Williams, and R. Bhaumik, "Detecting Profile Injection Attacks in Collaborative Recommender Systems," the 8th IEEE International Conference on E-Commerce Technology and the 3rd IEEE International Conference on Enterprise Computing, E-Commerce, and E-Services, CEC/EEE, 2006.

[12] B. Mehta and W. Nejdl, "Unsupervised strategies for shilling detection and robust collaborative filtering," User Modeling and User-Adapted Interaction, vol. 19, no. 1-2, Springer, February 2009, pp. 65-97.

[13] C. A. Williams, B. Mobasher, and R. Burke, "Defending recommender systems: detection of profile injection attacks," Service-Oriented Computing and Applications, Springer-Verlag, London, August 2007, pp. 157-170.

[14] K. Bryan, M. O'Mahony, and P. Cunningham, "Unsupervised Retrieval of Attack Profiles in Collaborative Recommender Systems," in ACM Conference on Recommender Systems, RecSys 2008, Lausanne, Switzerland, October 2008, pp. 155-162.

[15] S. Zhang, A. Chakrabarti, J. Ford, and F. Makedon, "Attack detection in time series for recommender systems," in ACM SIGKDD International Conference on Knowledge Discovery and Data Mining, KDD 2006, Philadelphia, Pennsylvania, USA, August 2006, pp. 809-814.

[16] M. Gao, Q. Yuan, B. Ling, and Q. Xiong, "Detection of abnormal item based on time intervals for recommender systems," The Scientific World Journal, Hindawi Publishing Corporation, 2014.

[17] Z. Wu, J. Cao, B. Mao, and Y. Wang, "Semi-SAD: applying semi-supervised learning to shilling attack detection," in ACM Conference on Recommender Systems, RecSys 2011, Chicago, Illinois, USA, October 2011, pp. 289-292.

[18] X. Shen, and R. Wu, "Discussion on t-Distribution and Its Application," Statistical and Application, Hans, December 2015, pp. 319-334.